\documentclass[12pt]{article}
\usepackage{amsmath}
\usepackage{amssymb}
\usepackage{citesort}
\usepackage{graphicx}
\usepackage[nomarkers]{endfloat}
\usepackage{pstricks}
\numberwithin{equation}{section}
\begin{document}

\setcounter{page}{0}
\thispagestyle{empty}

\begin{flushright}
{\small BARI-TH 384/00}
\end{flushright}

\vspace*{2.5cm}

\begin{center}
{\large\bf Influence of the magnetic field on the
\\[0.4cm]fermion scattering off bubble and kink walls }
\end{center}

\vspace*{2cm}

\renewcommand{\thefootnote}{\fnsymbol{footnote}}

\begin{center}
{ P. Cea$^{1,2,}$\protect\footnote{Electronic address: {\tt
Cea@bari.infn.it}},
G.~L. Fogli$^{1,2,}$\protect\footnote{Electronic address: {\tt
Fogli@bari.infn.it}} and
L. Tedesco$^{1,2,}$\protect\footnote{Electronic address: {\tt
Tedesco@bari.infn.it}} \\[0.5cm] $^1${\em Dipartimento di Fisica,
Universit\`a di Bari, I-70126 Bari, Italy}\\[0.3cm] $^2${\em INFN
- Sezione di Bari, I-70126 Bari, Italy} }
\end{center}

\vspace*{0.5cm}

\begin{center}
{
May, 2000 }
\end{center}

\vspace*{1.0cm}

\renewcommand{\abstractname}{\normalsize Abstract}
\begin{abstract}
We investigate the scattering of fermions off domain walls at the
electroweak phase transition in presence of a magnetic field. We
consider both the bubble wall and the kink domain wall. We derive
and solve the Dirac equation for fermions with momentum
perpendicular to the walls, and compute the transmission and
reflection coefficients. In the case of kink domain wall, we
briefly discuss the zero mode solutions localized on the wall. The
possibile role of the magnetic field for the electroweak
baryogenesis is also discussed.
\end{abstract}
%
\vfill
\newpage
\renewcommand{\thesection}{\normalsize{\Roman{section}.}}
\section{\normalsize{INTRODUCTION}}
\renewcommand{\thesection}{\arabic{section}}
Recently, a considerable amount of theoretical work, based on the
Standard Model of electroweak interactions and its extensions, has
been done on the hypothesis that the production of the observed
baryon asymmetry may be  generated at the primordial electroweak
phase transition~\cite{Trodden:1999}. Indeed, it has been
recognized that the baryon number is violated in the Standard
Model through the axial anomaly~\cite{tHooft:1976}. Thus, it
widely believed that the baryon asymmetry may originate in the
Standard Model if the electroweak phase transition is of first or
weakly first order. \\
In general, the problem of determining the nature of the
electroweak phase transition requires a non perturbative approach.
As a matter of fact, in lattice simulations it
turns out that the electroweak phase transition is first order for
Higgs masses less than $ 72.1 \pm 1.4 \;\; {\rm {GeV}} $, while for larger
Higgs masses only a rapid crossover is expected~\cite{Fodor:2000}.
Thus, if we take into account the lower bound
recently established by the LEP Collaborations~\cite{ALEPH:1999}:
\begin{equation}
\label{bound}
M_H \; > \; 95.2 \; {\rm {GeV}} \; \;  {\rm {at}} \;  
\; 95 \;  \% \;  \; {\rm {C.L.}} \;
,
\end{equation}
then, we are led to conclude  that in the Standard Model 
there is no first order
phase transition. However, a first order phase transition can be
obtained with an extension of the Higgs
sector of the Standard Model~\cite{Shaposhnikov:1986}, or in the
minimal supersymmetric Standard Model~\cite{Comelli:1994}. 
This leaves open the door for a first order phase transition.
\\
On the other hand, even in a
perfectly homogeneous continuous phase transition defects will
form, if the transition proceeds faster than the order parameter
of the broken symmetry phase is able to relax ~\cite{Kibble:1976}. 
In such a non
equilibrium transition, the new low temperature phase starts to
form, due to quantum fluctuations, simultaneously and
independently in many parts of the system. Subsequently, these
regions grow to form the new broken phase. When different causally
disconnected regions meet, the order parameter does not necessarily
match and a domain structure is formed. Thus, in the Standard
Model the neutral Higgs field is expected to become organised into
domains, in each of which the field has a constant sign. As a
consequence, the defects are domain walls across which the field
flips from one minimum to the other. \\
The dynamics of the domain walls is governed by the surface
tension $\sigma$. In particular, it was pointed out by Zel'dovich,
Kobazarev and Okun~\cite{Zel'dovich:1974} that the gravitational
effects of just one such wall stretched across the  universe would
introduce a large anisotropy into the relic blackbody radiation.
For this reason, the existence of such walls was ruled out.
However,  it has been suggested recently~\cite{Cea:1999}, that the
effective surface tension of the domain walls can be made
vanishingly small, due to a peculiar magnetic condensation induced
by the fermion zero modes localized on the wall. As a consequence,
the domain wall acquires a non zero magnetic field perpendicular
to the wall and becomes invisible as for as  gravitational
effects are concerned. 
In particular, such ferromagnetic walls have been proposed as 
possible sources of the primordial cosmological magnetic
field~\cite{Cea:1999}.
On the other hand, even for the bubble walls it has been
suggested~\cite{Vachaspati:1991} that strong magnetic fields may
be produced as a consequence of non vanishing spatial gradients of
the classical value of the Higgs field. \\
One of the ingredients which enters in the dynamical generation of
the baryon asymmetry at the electroweak transition is the
estimate of the transmission and reflection coefficients for
the fermion scattering off the walls. 
In other words, it is interesting to
see how the magnetic field localized at the wall modifies these
coefficients.  Aim of this paper is to evaluate the reflection
and transmission coefficients for the scattering of Dirac fermions
off electroweak walls in presence of a magnetic field.
We expect that, in general, the magnetic penetration length is much
greater than the wall thickness. This means that fermions which
scatter on the wall feel an almost constant magnetic field over a
spatial region much greater than the wall thickness. So that we can
assume that the magnetic field is constant. Moreover, 
limit our interest to the case of 
consider magnetic fields  perpendicular to the wall 
and assume the momentum of 
Dirac fermions  asymptotically 
perpendicular to the wall surface. This corresponds to neglect the
motion parallel to the wall surface. As we shall see, this 
approximation allows to avoid inessential
technicalities. The full calculation will be presented in a future
work. \\
The plan of the paper is as follow. In Section~2 we discuss the
case of bubble wall. We solve the Dirac equation in presence of a
planar bubble wall with a constant magnetic field perpendicular to
the wall. We evaluate the reflection and transmission coefficients
and compare them with the known results without the magnetic field.
Section~3 is devoted to the case of planar domain walls. In this
case we discuss the solutions localized on the wall. Finally, some
concluding remarks are presented in Section~4. \\
\renewcommand{\thesection}{\normalsize{\Roman{section}.}}
\section{\normalsize{BUBBLE WALL}}
\renewcommand{\thesection}{\arabic{section}}
In a first order phase transition the conversion from one phase to
another occurs through nucleation of the true phase in the false
phase. If the electroweak phase transition is of first
order~\cite{Kirzhnitz:1972}, the bubble of the true phase expand
and eventually fills the entire volume. At the bubble wall the
matter is out of equilibrium and it is here that the baryon
asymmetry might be produced. To understand the spontaneous
generation of the baryon asymmetry we need to investigate the
effects of the bubble wall on the propagation of fermions.
Following Ref.~\cite{Ayala:1994}, we simplify the problem by
assuming the bubble, which is at rest, as a planar interface
separating the symmetric phase (outside the bubble) from the
broken phase (inside the bubble). Fermions passing through the
bubble acquire mass, generated by the Yukawa coupling, which is
proportional to vacuum expectation value of the Higgs field. \\
Assuming the bubble wall thickness  $\Delta = 1$, in our
approximation the wall profile reads:
\begin{equation}
\label{eq2.1}
\varphi(x_3)\; = \; \frac {v} {2} \; \; [1 \; \;  + \; \tanh(x_3)
\;] \; = \; v \; g(x_3) \; ,
\end{equation}
where $x_3$ is the direction normal to the wall surface, and $v$
is the vacuum expectation value of the Higgs field. \\
We are interested in the Dirac equation  in the bubble wall
background and in presence of the electromagnetic field
$A_\mu(x)$:
\begin{equation}
\label{eq2.2}
[i \; \gamma^{\mu} \; \partial_{\mu} - e \; \gamma^{\mu} \;
A_{\mu} - g_Y \; \varphi(x_3) ] \; \Psi = 0 \; ,
\end{equation}
where $e$ and $g_Y$ are  the  electric  and
Yukawa coupling respectively. To solve Eq.~(\ref{eq2.2}) we assume:
\begin{equation}
\label{eq2.3}
\Psi \;= \; [i \; \gamma^{\mu} \; \partial_{\mu} - e \;
\gamma^{\mu} \; A_{\mu} + g_Y \; \varphi(x_3) ] \Phi \; .
\end{equation}
Inserting into Eq.~(\ref{eq2.2}) we readily obtain:
\begin{equation}
\label{eq2.4}
[-\partial^2 - 2 i e A_\mu \partial^{\mu} + e^2 A_{\mu} A^{\mu} -
\xi^2 g^2(x_3) +i \xi \gamma^{\mu} \partial_{\mu} g(x_3) -  \frac
{1} {2} \, \sigma^{\mu \nu} F_{\mu \nu} ] \Phi \; = \; 0 \; ,
\end{equation}
where $2 \xi$ is the mass which the fermion acquires in the broken
phase. In the case of constant magnetic field directed along the
third spatial direction we have $ F_{1 2}  = B$. It is easy to see
that the solutions of Eq.~(\ref{eq2.4}) factorize as:
\begin{equation}
\label{eq2.5}
 \Phi(x_1,x_2,x_3,t) \; = \; \zeta(x_1,x_2)\; \phi(x_3,t) \; ,
\end{equation}
where $\zeta$ is a scalar function which describes the motion
transverse to the bubble wall. In this paper we consider particles
with momentum perpendicular to the wall. Then Eq.~(\ref{eq2.4})
reduces to:
\begin{equation}
\label{eq2.6}
 \left( -\partial_t^2  -\partial_3^2 - \xi^2 g^2(x_3) +i \xi \gamma^{3}
\partial_{3} g(x_3) -  e \, \sigma^{1 2} F_{1 2} \right)
 \phi(x_3,t) \; = \; 0
\; .
\end{equation}
Putting $\phi(x_3,t)=e^{\mp i E t} \omega (x_3) $, corresponding
to positive and negative energy solutions respectively, we get:
\begin{equation}
\label{eq2.7}
 \left( \frac {d^2} {d x_3^2} + i \gamma^3 \frac {d
g} {d x_3} - \xi^2 g^2(x_3) + E^2 + i e B \gamma^1 \gamma^2
\right) \omega(x_3) =0 \, .
\end{equation}
In order to solve  Eq.~(\ref{eq2.7}), we expand $\omega(x_3)$ in
terms of the spinors $u_{\pm}^s$  eigenstates of $\gamma^3$:
\begin{equation}
\label{eq2.8}
\omega(z) = \phi_+^1 u_+^1 + \phi_-^1 u_-^1 +
\phi_+^2 u_+^2 + \phi_-^2 u_-^2 \; .
\end{equation}
Using the standard representation for the Dirac
matrices~\cite{Bjorken:1962} we find:
\begin{eqnarray}
\label{eq2.9}
 u_{\pm}^1 &=& \left(
\begin{array}{c}
1 \\ 0 \\ \pm i\\ 0
\end{array}
\right)\ ,\\
\label{eq2.10}
 u_{\pm}^2 &=& \left(
\begin{array}{c}
0 \\ 1 \\ 0\\ \mp i
\end{array}
\right)\ \; .
\end{eqnarray}
It is straightforward to check that:
\begin{eqnarray}
\label{eq2.11}
\gamma^3 u_{\pm}^{1,2} &=& \pm i u_{\pm}^{1,2}
\nonumber
\\
\gamma^0 u_{\pm}^{1,2}&=& + u_{\mp}^{1,2} \nonumber
\\
\gamma^1 \gamma^2 u_{\pm}^1 &=& -i u_{\pm}^1 \nonumber
\\
\gamma^1 \gamma^2 u_{\pm}^2 &=& +i u_{\pm}^2 \, .
\end{eqnarray}
Taking into account
Eqs.~(\ref{eq2.8}),(\ref{eq2.9}),(\ref{eq2.10}), (\ref{eq2.11}) we
obtain the equations:
\begin{equation}
\label{eq2.12} \left( \frac {d^2} {d x_3^2}\mp\xi\frac {d g} {d z}
- \xi^2  g^2(x_3) + E^2 + eB \right) \phi_{\pm}^1=0\, ,
\end{equation}
\\
\begin{equation}
\label{eq2.13} \left( \frac {d^2} {d x_3^2}\mp\xi\frac {d g} {d
x_3}-\xi^2 g^2(x_3) + E^2 -eB \right) \phi_{\pm}^2=0 \; .
\end{equation}
As, expected the magnetic field introduces an explicit spin
dependence in the equations. Note that, in our approximation the
spin projection on the third spatial axis coincides with the
helicity.
\\
In order to solve the differential equations (\ref{eq2.12}),  
(\ref{eq2.13}), following
Ref.~\cite{Ayala:1994} we introduce the new variable:
\begin{equation}
\label{eq2.14}
y= \frac {1} {1+ e^{2 x_3}} .
\end{equation}
For definiteness, let us consider Eq.~(\ref{eq2.12}). We have:
\begin{equation}
\label{eq2.15}
\left[ \frac {d^2} {d y^2} + \frac {1-2y} {y(1-y)}
\frac {d} {d y} + \frac {E^2 + eB \mp 4 \xi y(1-y) -4
\xi^2(1-y)^2} {4 y^2 (1-y)^2}  \right] \phi_{\pm}^1=0 \; .
\end{equation}
By examining the behaviour of the differential equation near the
singular points, we get:
\begin{equation}
\label{eq2.15bis}
\phi_{\pm}^1(x_3) \; = \; y^{\alpha^1} \; ( 1 \; - \; y)^{\beta^1}
\;
 \chi_{\pm}^1(y) \; ,
\end{equation}
where $ \chi_{\pm}^1(y)$ is regular near the singularities. A
standard calculation gives:
\begin{eqnarray}
\label{eq2.16}
\alpha^1 &=&\frac{i}{2}\sqrt{E^2+eB-4 \xi^2}\, ,
\nonumber
\\
\beta^1 &=& \frac {i} {2} \sqrt{E^2+eB} \; .
\end{eqnarray}
Inserting Eq.~(\ref{eq2.15bis}) into Eq.~(\ref{eq2.15}) one
sees that $ \chi_{\pm}^1(y)$ satisfies the hypergeometric
equation~\cite{Gradshteyn:1963} with parameters:
\begin{eqnarray}
\label{eq2.17}
a_{\mp}^1 &=& \alpha^1+\beta^1 + \frac {1} {2} -
\left|\xi \mp \frac {1} {2} \right|, \nonumber
\\
b_{\mp}^1 &=& \alpha^1 + \beta^1 + \frac {1} {2} + \left|\xi \mp
\frac {1} {2} \right|, \nonumber
\\
c^1 &=& 2 \alpha^1 +1 \; .
\end{eqnarray}
As  well known, the general solution of the hypergeometric
equation is a combination of the two independent solutions:
\begin{eqnarray}
\label{eq2.18}
 _{2}F_{1}(a_{\mp},b_{\mp},c,y) \, ,
\nonumber
\\
y^{1-c} \; _{2}F_{1}(a_{\mp}+1-c,b_{\mp}+1-c,2-c;y) \; .
\end{eqnarray}
Therefore:
\begin{eqnarray}
\label{eq2.19}
 \phi_{\pm}^1 &=& A^1_{\pm} \,  y^{\alpha^1} (1-y)^{\beta^1} \,
 _2F_1(a_{\mp}^1,b_{\mp}^1,c^1,y) + \nonumber
\\
&& +  B_{\pm}^1 \, y^{-\alpha^1} (1-y)^{\beta^1} \,
 _2F_1(a_{\mp}^1+1-c^1,b_{\mp}^1+1-c^1,2-c^1;y) \; ,
\end{eqnarray}
where $A_{\pm}^1$ and $B_{\pm}^1$ are normalization constants.
For simplicity we define:
\begin{eqnarray}
\label{eq2.19bis}
\phi_{\pm}^{(-\alpha^1)} &=& y^{\alpha^1}
(1-y)^{\beta^1} \, _2F_1(a_{\mp}^1,b_{\mp}^1,c^1;y)\; , \nonumber
\\
\phi_{\pm}^{(+\alpha^1)} &=& y^{-\alpha^1} (1-y)^{\beta^1} \,
  _2F_1(a_{\mp}^1+1-c^1,b_{\mp}^1+1-c^1,2-c^1;y) \; .
\end{eqnarray}
The  fermionic wave function can be obtained from
Eq.~(\ref{eq2.3}), which in our approximation reduces to:
\begin{equation}
\label{eq2.20}
\Psi^1(x_3,t)=[- i \gamma^0 \partial_0+i\gamma^3
\partial_3
             + \xi \, g(x_3) ] \, e^{\mp i E t} \,  \phi^1_{+} \, u^1_{+} \; .
\end{equation}
A calculation similar to that performed in Ref.~\cite{Ayala:1994} gives:
\begin{eqnarray}
\label{eq2.21}
\Psi^1(x_3)&=& A^1 \, [\epsilon_r \, {\phi_+^1}^{(-\alpha^1)} \,
u_-^1
            +2(\xi+\alpha^1) {\phi_-^1}^{(-\alpha^1)} \, u_+^1]+
\nonumber
\\
& &            B^1 \, [\epsilon_r \, {\phi_+^1}^{(+\alpha^1)}
u_-^1
            +2(\xi-\alpha^1) {\phi_-^1}^{(+\alpha^1)} \, u_+^1]
\end{eqnarray}
where $\epsilon_r=E$ for $r=1$ and  $\epsilon_r=-E$ for $r=2$.
\\
We are interested in fermions incident from the symmetric phase region
( $x_3 \rightarrow - \infty$), reflected in part by the
bubble wall  and transmitted in the broken phase region ($ x_3 \rightarrow +
\infty$). We need then the asymptotic forms of
$\phi_{\pm}^{(-\alpha^1)}$ and $\phi_{\pm}^{(+\alpha^1)}$:
\begin{equation}
\label{eq2.22}
\lim_{x \rightarrow +\infty} y^{\pm \alpha^1}
(1-y^{\beta^1}) = \exp(\mp 2 \alpha^1 x_3)\, ,
\end{equation}
\begin{equation}
\label{eq2.23}
\lim_{x \rightarrow -\infty} y^{\alpha^1} (1-y^{\pm
\beta^1}) = \exp(\pm 2 \beta^1 x_3)\, .
\end{equation}
After some algebra we obtain the transmitted, incident and
reflected  wave functions:
\begin{equation}
\label{eq2.24}
(\Psi^1)^{trans}=A \, [\epsilon_r u^1_- + 2 (\xi - \alpha^1) u_+]
\exp( 2 \alpha^1 x)\, ,
\end{equation}
\\
\begin{equation}
\label{eq2.25}
(\Psi^1)^{inc}=A \frac {\Gamma(1-2 \alpha^1) \Gamma(-2 \beta^1)}
{\Gamma(- \alpha^1 -\beta^1 + \xi) \Gamma(- \alpha^1 -\beta^1 -
\xi)} \left[ \frac {\epsilon_r u_-^1} {- \alpha^1 -\beta^1 - \xi}
+ \frac {2 (\xi - \alpha^1) u_+^1} {- \alpha^1 -\beta^1 + \xi}
\right] e^{2 \beta^1 x_3} \, ,
\end{equation}
\\
\begin{equation}
\label{eq2.26}
(\Psi^1)^{ref}=A \frac {\Gamma(1-2 \alpha^1) \Gamma(2 \beta^1)}
{\Gamma(- \alpha^1 +\beta^1 + \xi) \Gamma(- \alpha^1 +\beta^1 -
\xi)} \left[ \frac { \epsilon_r u_-^1} {- \alpha^1 +\beta^1 - \xi}
+ \frac {2 (\xi - \alpha^1) u_+^1} {- \alpha^1 +\beta^1 + \xi}
\right] e^{- 2 \beta^1 x_3}.
\end{equation}
The relevant current turns out to be:
\begin{equation}
\label{eq2.27}
j^3_V={\bar {\psi}} \gamma_3 \psi \, .
\end{equation}
Taking into account
Eqs.~(\ref{eq2.24}),~(\ref{eq2.25}),~(\ref{eq2.26}), we get:
\begin{equation}
\label{eq2.28}
(j^3_V)^{1, inc}= 8i \epsilon_r \beta^1 |A|^2
\left|\frac {\Gamma (1+ 2 \alpha^1) \Gamma(2 \beta^1)} {\Gamma(
\alpha^1 + \beta^1 + \xi) + \Gamma( \alpha^1 + \beta^1 - \xi)}
\right|^2 \frac {1} {( \alpha^1 + \beta^1 + \xi) ( \alpha^1 +
\beta^1 - \xi)}\, ,
\end{equation}
\\
\begin{equation}
\label{eq2.29}
(j^3_V)^{1,trans}=- 8 i \epsilon_r \alpha^1 |A|^2 \, ,
\end{equation}
\\
\begin{equation}
\label{eq2.30}
(j^3_V)^{1, refl}= -8i \epsilon_r \beta^1 |A|^2
\left|\frac {\Gamma (1+ 2 \alpha^1) \Gamma(2 \beta^1)} {\Gamma(
-\alpha^1 + \beta^1 + \xi)  \Gamma( - \alpha^1 + \beta^1 - \xi)}
\right|^2 \frac {1} {( \alpha^1 - \beta^1 - \xi) ( \alpha^1 -
\beta^1 + \xi)}
\end{equation}
By means of the identity
\begin{equation}
\label{eq2.31}
\Gamma(z) \Gamma(-z) = \frac {- \pi} {z \sin (\pi
z)},
\end{equation}
we obtain the following reflection and transmission coefficients:
\begin{equation}
\label{eq2.32}
R^{1}= \frac {\sin[\pi(\alpha^1 - \beta^1 + \xi)] \, \sin[\pi(-
                   \alpha^1 + \beta^1 + \xi)]}
{\sin[\pi(- \alpha^1 - \beta^1 + \xi)]
\,  \sin[\pi(+ \alpha^1 + \beta^1 + \xi)]}
\end{equation}
\begin{equation}
\label{eq2.33}
T^{1}= \frac {\sin (2 \pi \alpha^1) \sin (2 \pi \beta^1)}
{\sin[\pi(- \alpha^1 - \beta^1 + \xi) \sin[\pi(+ \alpha^1 +
\beta^1 + \xi)]} \; ,
\end{equation}
where the superscript $1$ refers to the spin projection. \\
Note that the magnetic field influences the coefficients by means
the parameters $\alpha^1, \beta^1$. 
It is easily checked that $R^2$ and
$T^2$ are given by the same  Eqs.~(\ref{eq2.32}) and~(\ref{eq2.33}) with
$\alpha^1$ and $\beta^1$ replaced by:
\begin{eqnarray}
\label{eq2.34} \alpha^2 &=& \frac {i} {2} \sqrt{E^2-eB -4 \xi^2}\, ,
\nonumber
\\
\beta^2  &=& \frac {i} {2} \sqrt{E^2-eB} \; ,
\end{eqnarray}
which, as expected,  correspond to $B \longrightarrow -B$.
One can verify that $R + T = 1$ and that 
for $B=0$ $R^{1,2}$ and  $T^{1,2}$ agree with the
results of Ref.~\cite{Ayala:1994}. \\
In Figure~1 we compare the behaviour of $R^{1}$ and $R^{2}$ as a
function of the scaled energy $E/2 \xi$ with the zero
magnetic field reflection coefficient. We see that the magnetic
field is able to produce an asymmetry in the spin distribution,
but there is no particle-antiparticle asymmetry, which is relevant
in the electroweak baryogenesis. In particular, there is  total
reflection for fermions with antiparallel spin and energies $E^2
-4 \xi^2 \leq eB $.
\\
In Figure~2 we report the reflection coefficients at a given
energy as a function of $eB$. We see that by increasing the
magnetic field the reflection coefficient decreases for fermions
with parallel spin, and increases for antiparallel spin. Thus
the magnetic field enhances  the spin asymmetry in the scattering
off the bubble wall.
%
%
%
\renewcommand{\thesection}{\normalsize{\Roman{section}.}}
\section{\normalsize{KINK DOMAIN WALL}}
\renewcommand{\thesection}{\arabic{section}}
In this Section we consider the domain walls which are thought to
be formed in a continuous phase transition by the Kibble
mechanism~\cite{Kibble:1976}. Indeed, if the scalar field develops
a non vanishing vacuum expectation  value $\langle \phi \rangle =
\pm v$, then in general one may assume that
 there are regions with $\langle \phi \rangle = + v$ and
$\langle \phi \rangle = - v$. It is easy to see that the classical
equation of motion of the scalar field admits the solution
describing the transition layer between two adjacent regions with
different values of $\langle \phi \rangle$ (assuming the domain
wall thickness  $\Delta = 1$):
\begin{equation}
\label{eq3.1}
\varphi(x_3)\; = \; v  \;  \tanh(x_3) \; = \; v \; g(x_3) \; .
\end{equation}
The solution (\ref{eq3.1}) is known as kink. In the same
approximation of the previous Section we must solve the Dirac
equation~(\ref{eq2.7}) with $g(x_3)$ given by
Eq.~(\ref{eq3.1}), $\xi$ being the fermion mass in the broken
phase. Again we write the spinor  $\phi(x_3)$ in the basis
(\ref{eq2.9})  and get four different equations. For
definiteness we focus on the equation for  $\phi(x_3)=
\phi^1_+(x_3) u^1_+$. We obtain:
\begin{equation}
\label{eq3.2}
\left( \frac {d^2} {d x_3^2} - \xi \frac {d g} {d x_3} - \xi^2
g^2(x_3) + E^2 + eB \right) \phi_{+}^1 = 0 \; .
\end{equation}
Following the approach of the previous Section, we find:
\begin{equation}
\label{eq3.4}
 \phi^1_+(x_3) \; = \; y^{\alpha^1} \; ( 1 \; - \; y)^{\alpha^1} \;
 \chi^1_+(y) \; ,
\end{equation}
where $ \chi^1_+(y)$ is regular near the singularities, $y$ is
defined in Eq.~(\ref{eq2.14}) and:
\begin{equation}
\label{eq3.5}
 \alpha^1 \; = \; \frac{i}{2} \, \sqrt{E^2+eB - \xi^2}
 \; .
\end{equation}
Again we find that $ \chi^1_+(y)$ satisfies a hypergeometric
equation with parameters:
\begin{eqnarray}
\label{eq3.6}
 a &=& 2 \alpha^1 + \frac {1} {2} -
\left|\xi - \frac {1} {2} \right|, \nonumber
\\
b &=& 2 \alpha^1  + \frac {1} {2} + \left|\xi - \frac {1} {2}
\right|, \nonumber
\\
c &=& 2 \alpha^1 +1 \; .
\end{eqnarray}
After some calculations we obtain the general solution of the 
Dirac equation. We have:
\begin{eqnarray}
\label{eq3.7}
\Psi^1(x_3)&=& A^1 \,  y^{\alpha^1} (1-y)^{\alpha^1} \, \left[ \right.\, E
    \, _2F_1(2 \alpha^1 + \xi, 2 \alpha^1 - \xi + 1, 2 \alpha^1 + 1;y)\,
    u_-^1+
 \nonumber \\
 & &            + \,  (\xi+ 2 \alpha^1) \,
             _2F_1(2 \alpha^1 - \xi, 2 \alpha^1 + \xi + 1, 2 \alpha^1 + 1;y)
              \, u_+^1  \,\left. \right] +
\nonumber
\\
& &           +  B^1 \, y^{-\alpha^1} (1-y)^{\alpha^1} \, \left[ \right. \, E
               \,
  _2F_1( \xi, 1 - \xi, 1 - 2 \alpha^1 ;y) \, u_-^1 +
 \nonumber
 \\
 & &          + \, (\xi- 2 \alpha^1) \, _2F_1(-\xi, 1 + \xi, 1 - 2 \alpha^1 ;y)
             \, u_+^1  \, \left. \right] \; .
\end{eqnarray}
Note that in the present case both the regions $x_3 \rightarrow \pm
\infty$ correspond to the broken phase. The case of a fermion
incident from $x_3 \rightarrow - \infty$ is given by
Eq.~(\ref{eq3.7}) with $A^1 = 0$. Using the asymptotic expansion
of the hypergeometric functions~\cite{Gradshteyn:1963}, we obtain
the incident, reflected and transmitted wave functions:
\begin{equation}
\label{eq3.8}
(\Psi^1)^{inc}=B^1 \frac {\Gamma(1-2 \alpha^1) \Gamma(-2
\alpha^1)} {\Gamma(- 2 \alpha^1  + \xi) \Gamma(- 2 \alpha^1 -
\xi)} \left[ \frac { E \, u_-^1} {- 2 \alpha^1 - \xi} + u_+^1 \right
] e^{2 \alpha^1 x_3} \; ,
\end{equation}
\\
\begin{equation}
\label{eq3.9}
 (\Psi^1)^{ref}=B^1 \frac {\Gamma(1-2 \alpha^1)
\Gamma(2 \alpha^1)} {\Gamma( \xi ) \Gamma( - \xi )} \left[ \frac {
E \, u_-^1} { - \xi} + \frac {\xi - 2 \alpha^1) u_+^1} { \xi} \right]
 e^{-2 \alpha^1 x_3} \; ,
\end{equation}
\\
\begin{equation}
\label{eq3.10}
(\Psi^1)^{trans}=B^1 \, [ E u^1_- + (\xi - 2 \alpha^1) u^1_+]
 e^{2 \alpha^1 x_3} \; .
\end{equation}
Finally, the reflection and transmission coefficients are given
by:
\begin{equation}
\label{eq3.11}
R^{1}= \frac {\sin^2(\pi \xi) } {\sin[\pi(- 2 \alpha^1 + \xi)]
\sin[\pi(+ 2 \alpha^1 + \xi)]} \, ,
\end{equation}
\begin{equation}
\label{eq3.12}
T^{1}= \frac {\sin (2 \pi \alpha^1) \sin (- 2 \pi \alpha^1)}
{\sin[\pi(- 2 \alpha^1 + \xi)] \, \sin[\pi(+ 2 \alpha^1 + \xi)]} \; .
\end{equation}
$R^2$ and $T^2$ are easily obtained from the previous
equations by $B \rightarrow - B$. The dependence of the reflection
and transmission coefficients  on the magnetic field is quite
similar to the bubble wall case (see Fig.~3). \\
Let us conclude this Section by discussing the fermion states
localized on the wall. Indeed, it is known that the differential
equation:
\begin{equation}
\label{eq3.13}
 \left( \frac {d^2} {d x_3^2} + i \gamma^3 \frac {d
g} {d x_3} - \xi^2 g^2(x_3) + E^2 + i e B \gamma^1 \gamma^2
\right) \omega(x_3) = 0
\end{equation}
admits zero energy solutions localized on the wall in absence of
the magnetic field~\cite{Jackiw:1976}. For the scattering problem
we are discussing, these solutions are not relevant. However, in
presence of a non zero magnetic field it is easy to see that there
are localized states if:
\begin{equation}
\label{eq3.14}
 E^2 \pm eB \; = \; 0
\end{equation}
Thus, we see that there is a localized state:
\begin{equation}
\label{eq3.15}
 \omega(x_3) \; = \; e^{-\kappa(x_3)} \;  u_{-}^2 \; \; , \; \; \;
  \frac {d \kappa} {d x_3} \, = \, \xi g(x_3)
\end{equation}
when $ E^2 = eB $. In general, if we take into account the motion
parallel to the wall, it can be shown that there are localized
states for  $ E^2 =  2n \, eB \, , n =1,2, ... $ . \\
It is worthwhile to stress that the localized states, which are a
peculiar characteristic  of the domain walls, may play an
important role in the dynamics of the walls.
\renewcommand{\thesection}{\normalsize{\Roman{section}.}}
\section{\normalsize{CONCLUSIONS}}
\renewcommand{\thesection}{\arabic{section}}
In this paper we have investigated the effects of a constant
magnetic field on the scattering of fermions on planar walls. We
have discussed both the bubble wall, which is relevant for a first
order electroweak phase transition, and the kink domain wall. In
particular, we solved the Dirac equation for fermions with
momentum perpendicular to the walls, and computed the transmission
and reflection coefficients. As expected, we find that the
constant magnetic field induces a spin asymmetry in the fermion
reflections and transmissions. The results obtained are
interesting, but do not allow to produce directly asymmetries in
some local charges which is a prerequisite for the idea of
electroweak baryogenesis~\cite{Trodden:1999}. However,  before 
reaching any conclusions, one should
take into account that the magnetic field distorts  the fermion
thermal equilibrium distribution, so that it could in principle
induce an asymmetry between fermion and antifermion distributions.
For instance, if the magnetic field modifies the chemical
potential, then the fermion and antifermion
distributions are expected to be different. 
Such an analysis, however, requires
the study of the quantum Boltzmann transport
equation~\cite{Kadanoff:1962}.
\\
Let us conclude this paper by briefly discussing the role of the
domain walls in the electroweak baryogenesis. Indeed, the baryon
number dynamical generation induced by bubble walls has been
extensively discussed in the literature. Coversely, the
possible role of kink domain walls has never discussed before, due
to the fact that the existence of such walls was ruled out by the
Zel'dovich, Kobazarev and Okun argument~\cite{Zel'dovich:1974}. As
we argued before, ferromagnetic domain walls with vanishing
effective surface tension are an open possibility~\cite{Cea:1999}.
As a consequence, it is worthwhile to address ourself on possible
mechanisms to generate baryon number by domain walls. 
\\
As well
known, the baryon number violation in the electroweak theory takes
place through saddle-point configurations called sphalerons. In
general, the sphaleron mechanism is not exponentially suppressed
in regions where the vacuum expectation value of the Higgs field
is small. Thus, in the case of domain walls, the sphaleron mediated
baryon number generation can be active only in the small
transition layer of the wall. Apparently this fact seems to rule
out any relevance of the domain walls in the electroweak
baryogenesis. However, in Section~3 we find that the kink domains
in a constant magnetic field display fermion  states localized on
the wall.  For these trapped fermions the sphaleron could
efficiently converts any local charge asymmetry into a baryon
number asymmetry,  and these processes could be active for a
long time and spread over a large region of the  primordial
universe. If this is the case, we see that the actual value of the
baryon number asymmetry could be accounted for even with the tiny
amount of Standard Model CP violation. 
%

%
%
\begin{figure}[H]
\label{Fig1}
\begin{center}
\includegraphics[clip,width=0.85\textwidth]{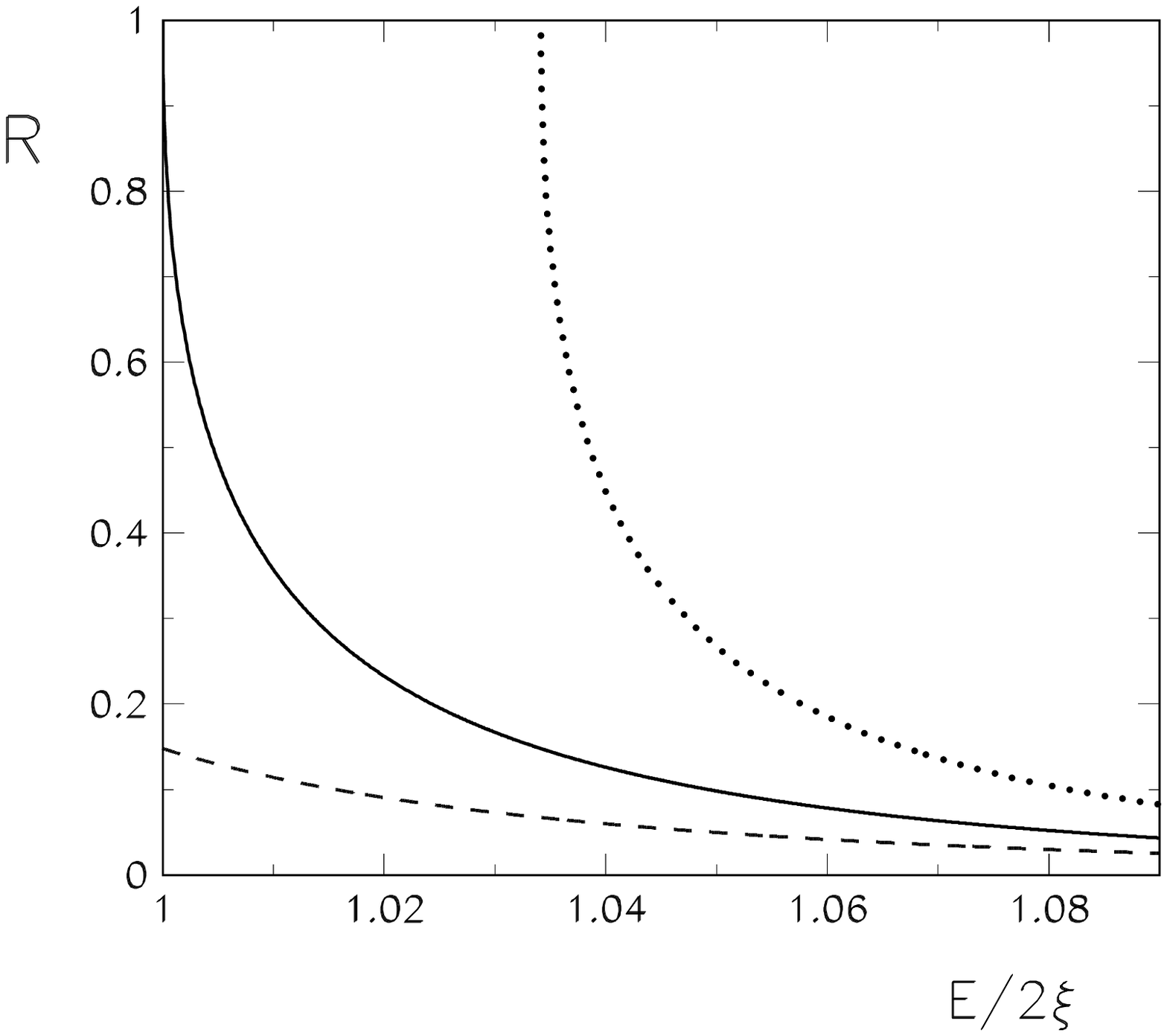}
\caption{Bubble wall reflection coefficients in terms of  the 
scaled energy $E/2 \xi$.
The dashed and dotted lines
correspond to $eB = 0.1$ and parallel and antiparallel spin respectively,
and are to be compared with the case $B = 0$ (solid line). }
\end{center}
\end{figure}
\begin{figure}[H]
\label{Fig2}
\begin{center}
\includegraphics[clip,width=0.85\textwidth]{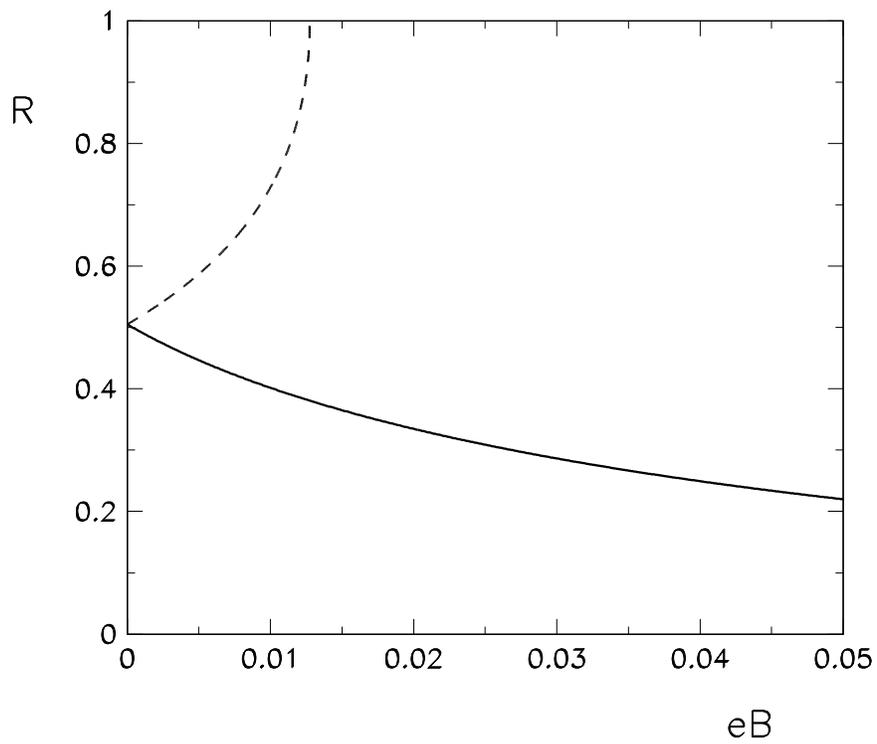}
\caption{Bubble wall reflection coefficients for parallel ( solid
line) and antiparallel (dashed line) spin as a function of $eB$
for $E=1.20$ and $\xi=0.6$.}
\end{center}
\end{figure}
\begin{figure}[H]
\label{Fig3}
\begin{center}
\includegraphics[clip,width=0.85\textwidth]{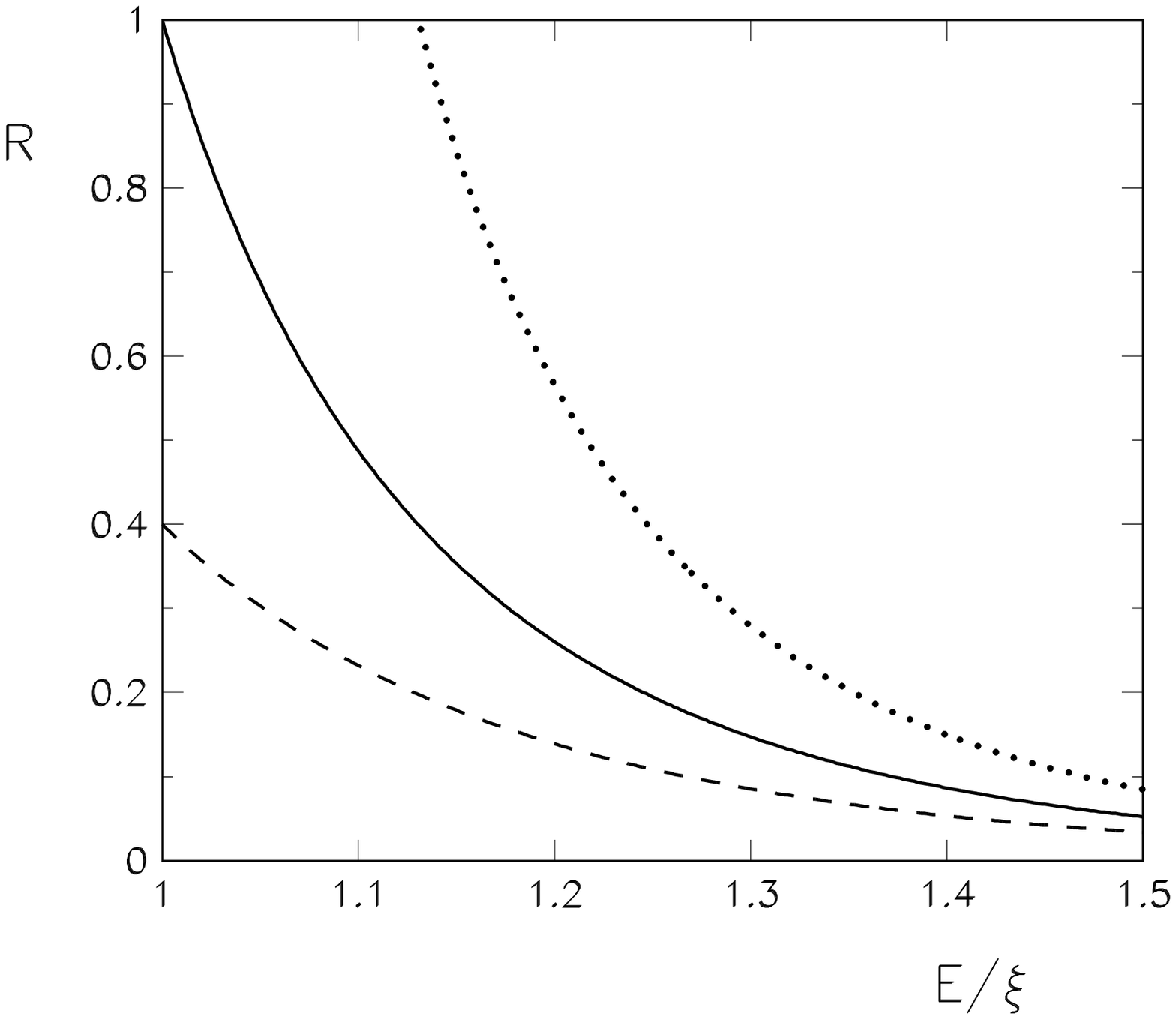}
\caption{ Reflection coefficients in the case of the domain wall.
Symbols as in Fig.~1. }
\end{center}
\end{figure}

\vfill
\newpage

\begin{thebibliography}{99}
\bibitem{Trodden:1999} For a recent review, see:
M.~Trodden, Rev. Mod. Phys. {\bf 71} (1999) 1463;
%
A.~Riotto, {\it Theories of Baryongenesis}, lectures delivered at
the {\it Summer School in High Energy Physics and Cosmology},
Miramare-Trieste, July 1998, hep-ph/9807454.
%
\bibitem{tHooft:1976}
G.~'t~Hooft, Phys. Rev. Lett. {\bf 37} (1976) 8;
%
V.~Kuzmin,V.~Rubakov, and M.~Shaposhnikov, Phys. Lett. {\bf B155}
(1985) 36.
%
\bibitem{Fodor:2000}For a recent review, see:
 Z.~Fodor,
Nucl.\ Phys.\ (Proc.\ Suppl.)\  {\bf 83-84} (2000) 121 .
%
\bibitem{ALEPH:1999} ALEPH, DELPHI, L3, OPAL and LEP Working Group
for Higgs Boson Searches, Technical Report No. ALEPH 99-081 CONF
99-052, International Europhysics Conference on High Energy
Physics, Tampere, Finland.
%
\bibitem{Shaposhnikov:1986}M.~E.~Shaposhnikov, JETP Lett. {\bf 44} (1986)
465; Nucl.\ Phys.\ {\bf B287} (1987) 757;  Nucl.\ Phys.\ {\bf
B299} (1988) 797; L.~McLerran, Phys. Rev. Lett. {\bf 62} (1989)
1075; N.~Turok and J.~Zadrozny, Nucl.\ Phys.\ {\bf B 358}
(1991)471; Nucl.\ Phys.\ {\bf B 369} (1992) 729;
 M.~Joyce, T.~Prokopec, and N.~Turok, Phys. Rev. {\bf D58}
(1995) 1234 ; J.~M.~Cline and A.~P.~Vischer, Phys. Rev. {\bf D 54}
(1996) 2451.
%
\bibitem{Comelli:1994}D.~Comelli, M.~Pietroni, and A.~Riotto, Nucl.\ Phys.\
{\bf B412} (1994) 441.
%
\bibitem{Kibble:1976}T.~W.~B.~Kibble, J.\ Phys.\
{\bf A9} (1976) 1387; Phys. Rep. {\bf 67} (1980) 183.
%
\bibitem{Zel'dovich:1974}Ya.~B.~Zeld'dovich, I.~Ya.~Kobzarev,
and L.~B.~Okun, JETP.\ {\bf 40} (1974) 1.
%
\bibitem{Cea:1999}P.~Cea and L.~Tedesco,  Phys.\ Lett.\
{\bf B450} (1999) 61.
%
\bibitem{Vachaspati:1991}T.~Vachaspati,  Phys.\ Lett.\
{\bf B265} (1991) 258. See also: D.~Grasso and A.~Riotto, Phys.\
Lett.\ {\bf B418} (1998) 258.
%
\bibitem{Kirzhnitz:1972}D.~A.~Kirzhnitz, JETP \ Lett. \
{\bf 15} (1972) 529; D.~A.~Kirzhnitz and A.~D.~Linde, Phys.\
Lett.\ {\bf 72B} (1972) 471.
%
\bibitem{Ayala:1994}A.~Ayala, J.~Jalilian-Marian, L.~McLerran and
A.~P.~Vischer,  Phys. Rev. {\bf D49} (1994) 5559.
%
%
\bibitem{Bjorken:1962}J.~D. Bjorken and S.~Drell,
 {\it Relativistic Quantum Fields}, McGraw Hill, New York~(1962).
%
%
\bibitem{Gradshteyn:1963}I.~S.~Gradshteyn and I.~M.~Ryzshik, {\it Table of
Integrals, Series and Products}, Academic Press, New York~(1963).
%
\bibitem{Jackiw:1976}R.~Jackiw and C.~Rebbi,  Phys. Rev. {\bf D13}
(1976) 3398; A.~T.~Niemi and G.~W.~Semenoff, Phys. Rep. {\bf C
135} (1986) 100.
%
\bibitem{Kadanoff:1962}L.~P.~Kadanoff and A.~Baym,  {\it Quantum Statistical
Mechanics}, W.~A.~Benjamin (1962); E.~M. Lifshitz and
L.~P.~Pitaevskii,  {\it Physical Kinetics}, Pergamon Press
(1979).
%
\end{thebibliography}
\end{document}